\documentclass[%
twocolumn,
superscriptaddress,
 amsmath,amssymb,
 aps,
]{revtex4-2}

\usepackage[english]{babel}
\usepackage{graphicx}
\usepackage{dcolumn}
\usepackage{bm}
\usepackage{comment}
\usepackage{xcolor}
\usepackage{soul}
\usepackage[normalem]{ulem}
\usepackage{siunitx}

\begin{document}

\preprint{APS/123-QED}

\newcommand{\sstate}{\ensuremath{|50\mathrm{s}\rangle}}
\newcommand{\pstate}{\ensuremath{|50\mathrm{p}\rangle}}
\newcommand{\dstate}{\ensuremath{|50\mathrm{d}\rangle}}

\title{Interfacing Rydberg atoms with a chip-based superconducting microwave resonator using an AC Stark shifted single-photon transition}

\author{L. L. Brown} 
\author{I. K. Bhangoo}
\author{S. D. Hogan}
\address{Department of Physics and Astronomy, University College London, Gower Street, London WC1E 6BT, UK}

\begin{abstract}
Helium atoms in the 1s50s\,$^3$S$_1$ Rydberg level have been resonantly coupled to the $2\pi\times11.722$~GHz second harmonic mode of a chip-based superconducting coplanar waveguide microwave resonator. To achieve this, the single-photon electric-dipole-allowed 1s50s\,$^3$S$_1\rightarrow$ 1s50p\,$^3$P$_J$ transition was tuned into resonance with the resonator mode through the AC Stark shift induced by a second strong $2\pi\times3.350$~GHz microwave dressing field. The effects of this dressing field, and residual uncanceled DC electric fields at the location of the atoms close to the superconducting chip surface were interpreted with support from Floquet calculations of the energy level structure of the Rydberg states. To observe appreciable population transfer in the \SI{1}{\micro\second} atom-resonator interaction time using this transition, which had an electric dipole moment of $1500\,e\,a_0$, pulsed microwave fields were injected into the resonator. From the photon occupation number in the resonator mode under these conditions, the single-photon Rabi frequency associated with the coupling of the atoms to the resonator was estimated to be $\sim2\pi\times100$~Hz. These results represent an important step toward operation of this Rydberg-atom--superconducting-circuit interface in the single-photon strong coupling regime.  
\end{abstract}

\maketitle

\section{Introduction}\label{sec:introduction}

The strong single-photon electric-dipole transitions that occur at microwave frequencies between highly-excited Rydberg states in atoms, are well suited to interfacing with solid-state qubits in superconducting circuits. At angular frequencies close to $2\pi\times10$~GHz, the electric dipole transition moments between these Rydberg states, and between the quantized energy-levels in superconducting qubits are similar, and lie in the range from $1\,000\,e\,a_0$ to $10\,000\,e\,a_0$, where $e$ is the elementary charge, and $a_0$ is the Bohr radius~\cite{schoelkopf08a}. Interfaces between Rydberg atoms and superconducting qubits, that exploit these strong electric dipole transitions, for example, by resonant coupling to a common mode of a superconducting coplanar waveguide (CPW) microwave resonator, therefore offer a scalable approach to networking neutral atom~\cite{Henriet2020quantumcomputing} and superconducting~\cite{RevModPhys.93.025005} quantum processors. 

Strong optical transitions between low-lying energy levels are employed in the process of laser photoexcitation of the atoms to Rydberg states in such a hybrid quantum system. However, superconducting qubits only exhibit transitions at microwave frequencies. Consequently, Rydberg-atom--superconducting-circuit interfaces also provide opportunities to use telecom-band optical photons -- which can propagate with low loss over long distances in optical fibers -- to prepare or manipulate the quantum states of superconducting qubits in an interaction mediated by the atoms~\cite{Petrosyan_2019,PhysRevLett.120.093201,petrosyan24a}. This type of scheme, that exploits optical excitation to the hybridized quantum states of a Rydberg atom strongly coupled to one or more superconducting qubits through a common mode of a CPW resonator, would, for example, enable the realization of long-range, fiber-coupled networks of superconducting processors.

To efficiently interface Rydberg atoms with superconducting circuits, so that the optical photons used for Rydberg state excitation may be exploited to manipulate the photon number state of a superconducting CPW resonator~\cite{PhysRevA.105.013707}, or in the longer term the quantum state of a superconducting qubit, resonant coupling is best achieved using strong single-photon microwave transitions between Rydberg states. In experiments reported up to now with helium (He) Rydberg atoms, coherent coupling to the third harmonic mode of a $\lambda/4$ superconducting CPW resonator at $2\pi\times19.556$~GHz was achieved using single-color, and two-color two-photon transitions between the 1s55s\,$^3$S$_1$ and 1s56s\,$^3$S$_1$ Rydberg levels~\cite{morganCouplingRydbergAtoms2020,walkerCavityenhancedRamseySpectroscopy2020,brown_demonstration_2024}. This work was performed with stray DC electric fields above the superconducting chip surface that approached zero~\cite{walkerElectrometrySingleResonator2022}. However, the use of a two-photon Rydberg-Rydberg transition meant that the effective electric dipole transition moment available for coupling to a single photon in the resonator, for example, in the case of the two-color two-photon scheme, was only $\sim50\,e\,a_0$.

Experiments have also been reported in which rubidium (Rb) Rydberg atoms were coherently coupled to a superconducting CPW resonator operating in a similar frequency range~\cite{kaiser_cavity-driven_2022}. However, the propensity for Rb to adsorb on the cryogenically cooled surfaces of the superconducting chip used in this work resulted in stray electric fields that were challenging to compensate~\cite{hattermann12a}. Consequently, it was necessary to apply a comparatively large DC offset electric field of $\sim3.6$~V/cm, at the location of the atoms above the CPW resonator, to dominate over these stray fields and allow the isolation of a pair of Rydberg levels with a small differential static electric dipole polarizability, which were suitable for coupling to it. The trade-off with using a strong offset electric field to overcome the effects of stray fields emanating from the superconducting chip surface was that the resulting single-photon electric dipole transition moments were reduced to $\sim30\,e\,a_0$.

With good control over stray electric fields close to the surface of a superconducting chip on which a CPW resonator is fabricated -- through the choice of atomic species to minimize surface adsorption and effects of stray electric fields from adsorbates, and the use of a $\lambda/4$ resonator geometry in which the center conductor is open at one end but grounded at the other to avoid charge build-up~\cite{walkerElectrometrySingleResonator2022} -- we demonstrate here the coupling of the single-photon transition between the $1\mathrm{s}50\mathrm{s}\,^3\mathrm{S}_1$ ($|50\mathrm{s}\rangle$) and $1\mathrm{s}50\mathrm{p}\,^3\mathrm{P}_J$ ($|50\mathrm{p}\rangle$) Rydberg levels in He to the $2\pi \times \SI{11.752}{GHz}$ second harmonic mode of a superconducting CPW resonator operated at 3.66~K. To achieve this while maintaining tunability at the interface without strongly polarizing the atoms and increasing their sensitivity to stray DC electric fields, a strong microwave dressing field was applied to AC Stark shift the $\pstate$ state. A similar approach has, for example, been used previously to control long-range interactions between pairs of Rydberg atoms~\cite{BohlouliZanjani07a}. The dressing field used in the experiments here was detuned from the $1\mathrm{s}50\mathrm{p}\,^3\mathrm{P}_J \rightarrow 1\mathrm{s}50\mathrm{d}\,^3\mathrm{D}_J$ ($|50\mathrm{p}\rangle \rightarrow |50\mathrm{d}\rangle$) transition and had an amplitude of $\sim370$~mV/cm. This allowed the $|50\mathrm{s}\rangle \rightarrow |50\mathrm{p}\rangle$ transition, which occurs at $\omega_{\mathrm{50s,50p}} = 2\pi \times \SI{12.149181}{GHz}$ in zero field, to be tuned into resonance with the resonator. In this strong microwave dressing field, the single-photon electric dipole transition moment for the $|50\mathrm{s}\rangle \rightarrow |50\mathrm{p}\rangle$ transition remained large, at $d \sim 1500\,e\,a_0$. Consequently, the microwave power in the resonator in the experiments could be reduced by more than three orders of magnitude from that used previously~\cite{brown_demonstration_2024}, moving this hybrid quantum system an important step closer to operation in the single-photon strong-coupling regime. 

In the following, an overview of the experimental apparatus is provided in Section~\ref{sec:expt}. In Section~\ref{sec:calibration} measurements made to characterize the superconducting CPW resonator, and stray DC electric fields at the atom-resonator interface are presented. This is followed in Section~\ref{sec:dressingscheme} by a description of the microwave dressing scheme employed to AC Stark shift the $|50\mathrm{s}\rangle \rightarrow |50\mathrm{p}\rangle$ transition into resonance with the resonator, and an overview of the Floquet calculations performed to support the interpretation of the experimental data. In Section~\ref{sec:results} the results of the experiments are presented before the outcomes of this work are discussed and conclusions drawn in Sections~\ref{sec:discuss} and \ref{sec:conc}, respectively. 

\begin{figure}[tb]
    \centering
    \includegraphics[width=0.46\textwidth]{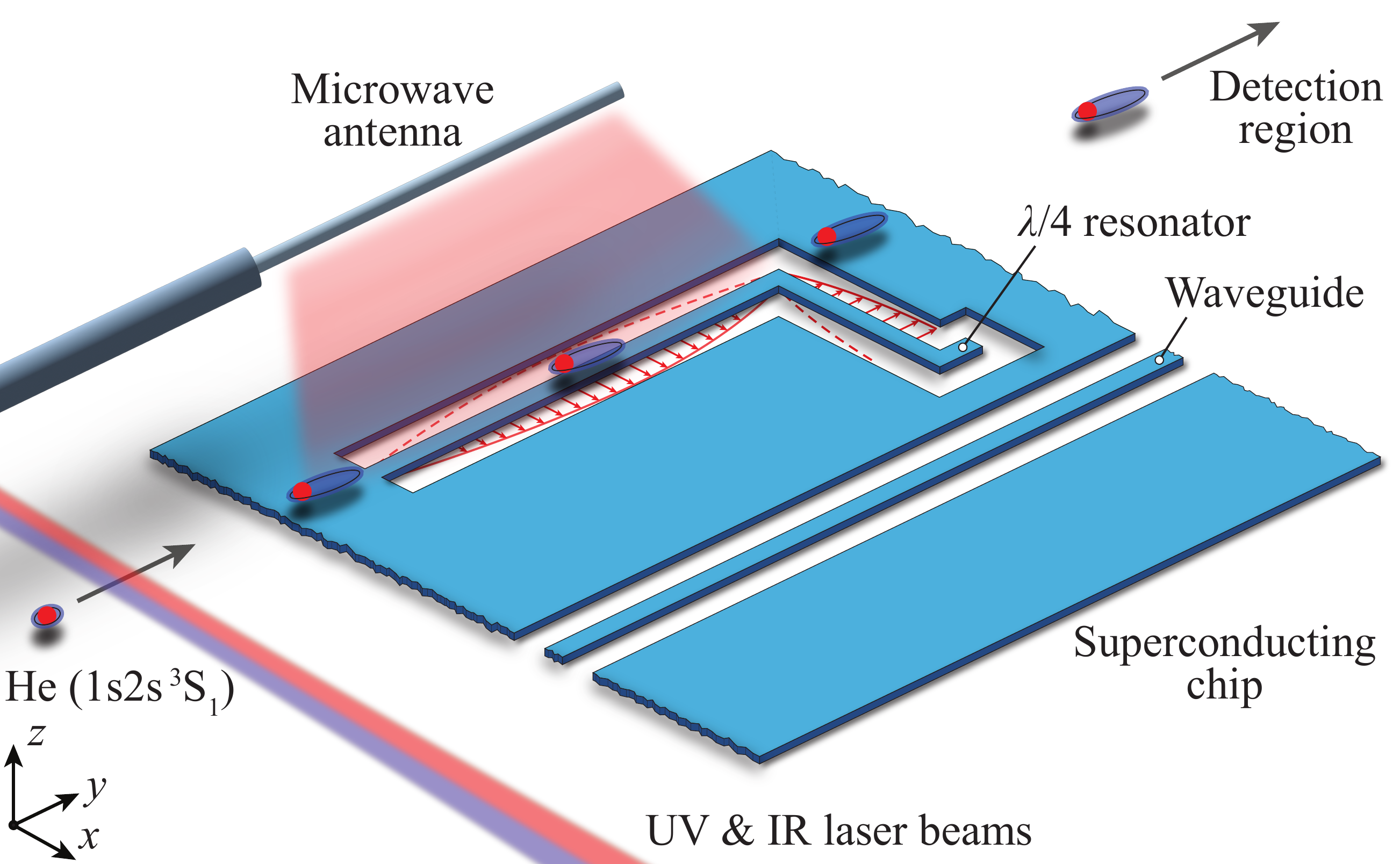}
    \caption{Schematic diagram of the Rydberg-atom--superconducting-circuit interface (not to scale). The microwave field distribution in the second harmonic mode of the resonator is indicated by the red arrows in the $xy$ plane. The strong microwave dressing field emanating from the antenna above the superconducting chip is represented by the shaded red band. See text for details.}
    \label{fig:expt}
\end{figure}

\section{Experiment}\label{sec:expt}

A schematic overview of the core cryogenic components of the apparatus used in the experiments is shown in Figure~\ref{fig:expt}. A pulsed supersonic beam of He atoms, traveling at a mean longitudinal speed of \SI{2000}{\meter\per\second}, was generated using a pulsed valve operated at a repetition rate of \SI{25}{Hz}. Atoms in this beam were prepared in the long-lived (\SI{7870}{\second} lifetime~\cite{hodgman_metastable_2009}) metastable 1s2s\,$^3$S$_1$ level in a DC electric discharge at the exit of the valve~\cite{halfmannSourceHighintensityPulsed2000}. The discharge was operated with a sharp metal anode (+230~V) positioned $\sim$1~mm from the valve opening, and seeded with electrons from a heated tungsten filament. Approximately 10~cm downstream from the valve, the beam was collimated by a 3-mm-diameter skimmer. Charged particles produced in the discharge were then removed by an electrostatic filter. The resulting neutral atom beam subsequently entered the cryogenic region of the apparatus through 4- and 3-mm-diameter apertures in heat shields maintained at $\sim$30~K and $<$4~K, respectively. 

Inside the sub-4~K core of the apparatus, the atomic beam was intersected at right angles by continuous-wave ultraviolet and infrared (IR) laser beams. These copropagating laser beams were focused to $\sim$\SI{100}{\micro\meter} full-width-at-half-maximum (FWHM) waists between a pair of parallel copper electrodes, and operated at wavelengths of 388.975 and 787.297~nm, respectively, to drive the resonance-enhanced two-color two-photon 1s2s\,$^3$S$_1$ $\rightarrow$ 1s3p\,$^3$P$_2$ $\rightarrow$ 1s50s\,$^3$S$_1$ excitation scheme~\cite{hoganLaserPhotoexcitationRydberg2018}. Short bunches of excited Rydberg atoms were generated by detuning the IR laser $\sim$~220~MHz below the field-free 1s3p\,$^3$P$_2$ $\rightarrow$ 1s50s\,$^3$S$_1$ transition frequency, and using a pulsed electric field of $\sim650$~mV/cm to Stark shift the Rydberg state into resonance with it for select periods of time. This electric field pulse was applied for \SI{3}{\micro\second}, resulting in the photoexcitation of 6-mm-long bunches of Rydberg atoms.

After excitation, the atoms traveled 25~mm to a $10~\mathrm{mm}\times10~\mathrm{mm}$ niobium nitride (NbN) superconducting chip (100-nm-thick NbN film on a silicon substrate), containing an L-shaped $\lambda$/4 CPW resonator (length 6.335~mm; center-conductor width \SI{20}{\micro\meter}; gap width \SI{10}{\micro\meter}), with one end grounded and the other open end capacitively coupled to a coplanar microwave waveguide~\cite{morganCouplingRydbergAtoms2020}. The CPW was connected to a microwave source outside the vacuum chamber. The coaxial cables used to achieve this were thermalized through attenuators on the 30~K and 4~K stages of the cryocooler. The total attenuation between the source and the CPW on the superconducting chip was 22.5~dB.

As seen in Figure~\ref{fig:expt}, the longer ($\sim5$~mm) straight section of the resonator was aligned with the propagation axis of the atomic beam. The Rydberg atoms, in a spatial distribution with a FWHM in the $z$ dimension of $\sim100~\mu$m, interacted primarily with the evanescent microwave field above this part of the resonator at a mean atom-surface distance of $\sim$\SI{300}{\micro\meter}~\cite{walkerElectrometrySingleResonator2022}. A split copper electrode was located $\sim10$~mm above the superconducting chip in the apparatus (not shown in Figure~\ref{fig:expt}, but visible in Figure~\ref{fig:fields}) to allow compensation of stray electric fields at the location of the atoms above the resonator. 

The resonance frequency of the resonator was coarsely controlled by setting the temperature, $T_{\mathrm{res}}$, of the NbN chip. In the experiments described here, the chip was operated at nominal temperatures of either $T_{\mathrm{res}} = 3.66$ or $5.00$~K, which resulted in resonance frequencies in the second harmonic mode of $\omega_2 \simeq 2\pi\times11.752$~GHz, or $2\pi\times11.710$~GHz, respectively. To couple the single-photon electric-dipole allowed $|50\mathrm{s}\rangle\rightarrow|50\mathrm{p}\rangle$ transition ($\omega_{50\text{s},50\text{p}} = 2\pi \times 12.149181$~GHz) to the second harmonic mode of the resonator, a strong microwave dressing field, $\omega_\text{dress}$ =  2$\pi \times$ 3.350~GHz, detuned by $\sim$2$\pi\times100$~MHz below the field-free $|50\mathrm{p}\rangle$ $\rightarrow$ $|50\mathrm{d}\rangle$ transition frequency ($|50\mathrm{d}\rangle \equiv$ 1s50d\,$^3$D$_J$) was generated above the superconducting chip. This field emanated from a microwave antenna located $\sim13$~mm above the chip surface as shown in Figure~\ref{fig:expt}, and it caused an AC Stark shift of the $\sstate\rightarrow\pstate$ transition toward lower frequencies -- as required to tune it into resonance with $\omega_2$. In each cycle of the experiment, the dressing field was applied as a 1-\SI{}{\micro\second}-duration pulse, \SI{12}{\micro\second} after Rydberg state photoexcitation. 

After passing over the superconducting chip the atoms moved into the detection region of the apparatus. There, 16~mm beyond the chip and \SI{25}{\micro\second} after initial laser photoexcitation, they were detected by state-selective pulsed electric field ionization. This was implemented by applying a slowly-rising negative potential to one of the electrodes in this region to ionize the atoms. The resulting electrons were then accelerated out of the cryogenic part of the apparatus to a microchannel plate detector operated at room temperature. The $|50\mathrm{s}\rangle$, $|50\mathrm{p}\rangle$ and $|50\mathrm{d}\rangle$ states involved in the experiments had calculated fluorescence lifetimes in the absence of external fields of $97$, $150$ and $56~\mu$s, respectively. These are all longer than the time between Rydberg-state laser photoexcitation and detection in each cycle of the experiment. However, to avoid complications in the interpretation of the experimental data resulting from these differences in lifetimes, or the differences in the electric field ionization efficiencies of these states, the interaction of the atoms with the resonator field was identified by selective detection and monitoring of the signal from the $\sstate$ state.

\begin{figure}
    \includegraphics[width=0.8\columnwidth]{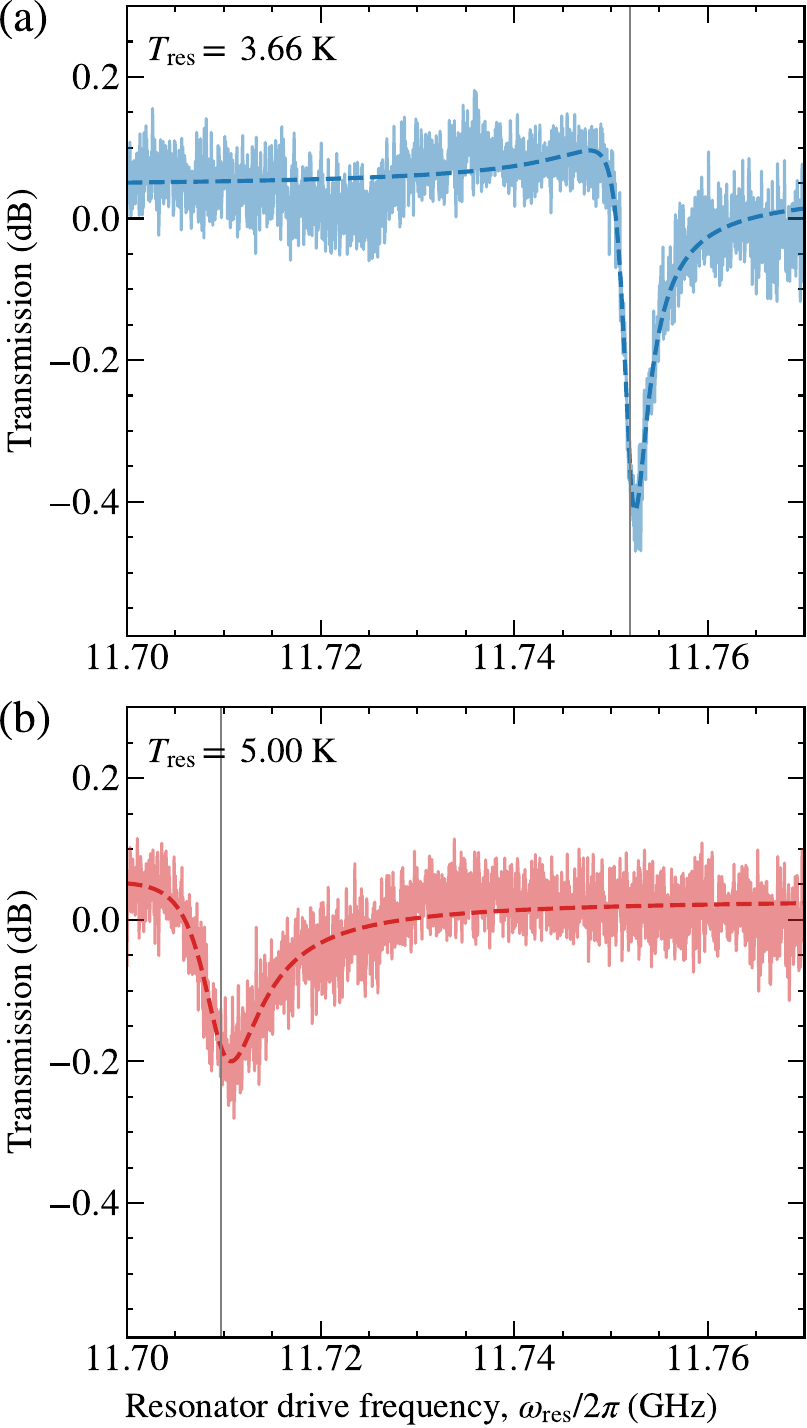}
    \caption{Microwave spectra measured in transmission through the superconducting circuit close to the second harmonic mode of the $\lambda/4$ CPW resonator recorded for (a) $T_\mathrm{res} =\SI{3.66}{\kelvin}$, and (b) $T_\mathrm{res} = \SI{5.00}{\kelvin}$. The dashed curves represent Fano lineshape functions fit to the experimental data using least squares methods. The continuous vertical lines in each panel indicate the resonance frequencies, $\omega_2$, determined from the fitting procedure.}
    \label{fig:transmission}
\end{figure}

\section{Superconducting CPW resonator and stray electric field characterization}\label{sec:calibration}

To characterize the CPW resonator, microwave spectra were recorded by transmission through the co-planar waveguide which was capacitively coupled to it on the superconducting chip. These spectra, obtained for $T_{\mathrm{res}} = 3.66$ and 5.00~K, and covering microwave frequencies close to $\omega_2$ are displayed in Figure~\ref{fig:transmission}(a) and~(b), respectively. In these data, a notch-type resonance is observed. This resonance lineshape is a consequence of interference between the microwave field transmitted directly through the waveguide, and that scattered from the resonator. To determine the resonance frequency, loaded quality factor and insertion loss from each spectrum, Fano functions of the form~\cite{fano1961},
\begin{eqnarray} 
S &=& A \frac{(q_{\mathrm{F}}+\epsilon)^2}{1+\epsilon^2} + C,\label{eq:fano func}
\end{eqnarray}
were fit to the data using least-squares methods. In this expression, the fit parameters were the Fano asymmetry parameter $q_{\mathrm{F}}$, the amplitude and offset $A$ and $C$, respectively, and the reduced energy $\epsilon = (\omega - \omega_{\mathrm{res}}) /(\Gamma/2)$ where $\omega_{\mathrm{res}}$ and $\Gamma$ are the resonance frequency and spectral FWHM, respectively. For $T_{\mathrm{res}} = 3.66$~K, this yielded a resonance frequency of $\omega_2 = 2\pi \times \SI{11.75198}{GHz}$, and a loaded quality factor, determined from the spectral FWHM, of $Q = 3960$. The insertion loss from the waveguide on-resonance was $L_\mathrm{ins}$ = \SI{26.67}{dB}. To perform experiments with the resonator off-resonance from the atomic transition frequency, the chip was also operated at the higher temperature of $T_\mathrm{res}$ = \SI{5.00}{\kelvin} [Figure~\ref{fig:transmission}(b)]. In this case the resonance frequency shifted to the lower value of $\omega_2 = 2\pi \times \SI{11.70969}{GHz}$,  the loaded quality factor reduced to $Q=1640$, and the insertion loss was $L_\mathrm{ins}$ = \SI{32.86}{dB}. The temperature-dependent shift of the resonator resonance frequency of $-2\pi\times\SI{42.29}{MHz}$ between Figure~\ref{fig:transmission}(a) and (b), is larger than the resonance width in either case. This was $\kappa = 2\pi\times \SI{2.97}{MHz}$ and $2\pi\times \SI{7.17}{MHz}$, respectively. The additional use of the phase of the microwave field in the circle-fit method~\cite{probstEfficientRobustAnalysis2015} to determine the resonator characteristics from the data in Figure~\ref{fig:transmission}, yielded values in agreement with those of the best fit Fano functions to within 10\%.

\begin{figure}
    \includegraphics[width=0.8\columnwidth]{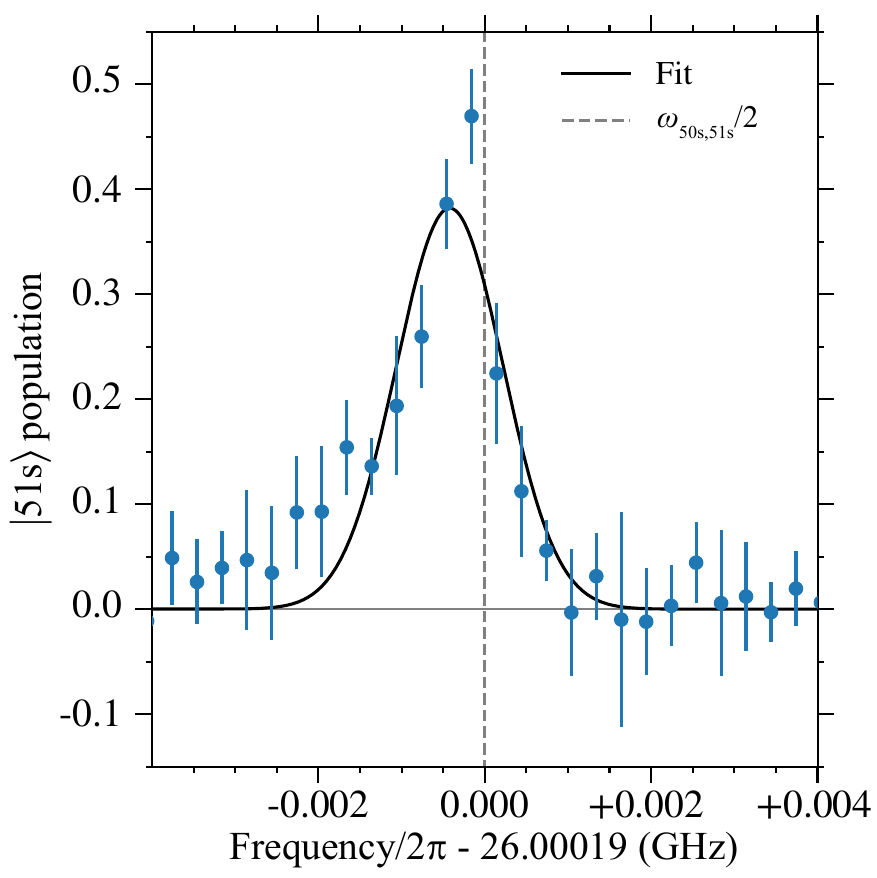}
    \caption{Microwave spectrum of the two-photon $|50\mathrm{s}\rangle \rightarrow |51\mathrm{s}\rangle$ transition recorded for He atoms located above the CPW resonator, but probed by a pulsed microwave field emanating from the antenna above the superconducting chip. The reference frequency of $\omega_{\mathrm{50s,51s}}/2 = 2\pi\times26.00019$~GHz on the horizontal axis is the field-free single-color two-photon $|50\mathrm{s}\rangle \rightarrow |51\mathrm{s}\rangle$ transition frequency.}
    \label{fig:fieldcancel}
\end{figure}

Stray electric fields in the region where the atoms interacted with the microwave field in the superconducting resonator were minimized in the experiments by applying pulsed offset electric potentials to a split electrode above the superconducting chip. The resulting compensation field was optimized, using the atoms as \textit{in situ} quantum sensors, by driving the antenna at frequencies close to the single-color two-photon $|50\mathrm{s}\rangle \rightarrow |51\mathrm{s}\rangle$ ($|51\mathrm{s}\rangle \equiv 1\mathrm{s}51\mathrm{s}\,^3\mathrm{S}_1$) transition frequency at $\omega_{\mathrm{50s,51s}}/2 = 2\pi\times \SI{26.001859}{GHz}$, and monitoring population transfer to the $|51\mathrm{s}\rangle$ state. An example spectrum recorded in this way, for a compensation potential of $+\SI{1.2}{V}$ is shown in Figure \ref{fig:fieldcancel}. When recording these data, the resonator and microwave dressing fields were off. The microwave pulses generated using the antenna had a duration of $\SI{1}{\micro\second}$, and were applied at the time in the experimental sequence when the atoms were located above the resonator. From this spectrum, a frequency shift of $-2\pi \times \SI{0.419}{MHz}$ from the field-free transition frequency is observed. Comparison of this shift with the calculated DC Stark shift of the two-photon transition, allowed the identification of a residual uncanceled electric field of $F_\mathrm{u}=104$~mV/cm at the location of the atoms.

\begin{figure}
    \centering
    \includegraphics[width=0.99\columnwidth]{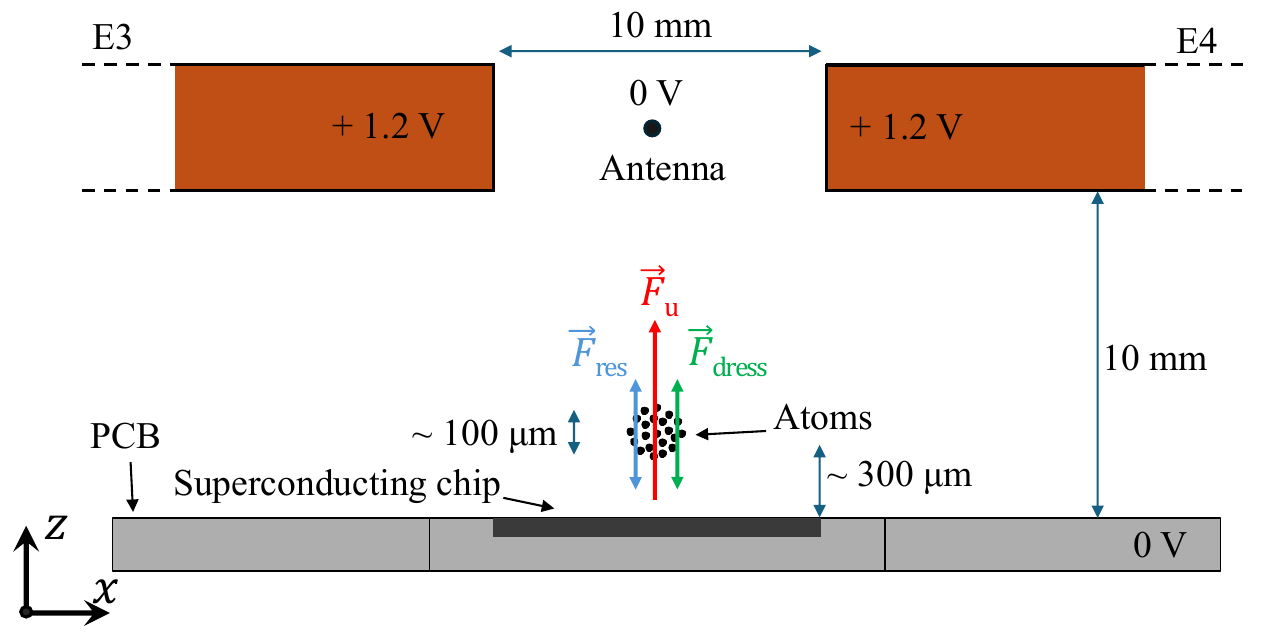}
    \caption{Diagram of the central region of the apparatus above the CPW resonator on the superconducting chip. At the position of the atoms when they interact with the resonator field, the linear polarization of the evanescent microwave field from the resonator, $\vec{F}_{\mathrm{res}}$, lies in the $z$ dimension and is parallel to the linear polarization of the microwave dressing field, $\vec{F}_{\mathrm{dress}}$, emanating from the antenna, and the dominant component of the residual uncanceled DC electric field, $\vec{F}_{\mathrm{u}}$.}
    \label{fig:fields}
\end{figure}

In previous work, with an apparatus similar to that in Figures~\ref{fig:expt} and~\ref{fig:fields}, but without the microwave antenna, between the split electrode components, E3 and E4, stray electric fields above the CPW resonator were compensated to $F_{\mathrm{u}} = 27$~mV/cm~\cite{walkerElectrometrySingleResonator2022}. The introduction of the antenna to generate strong microwave dressing fields above the superconducting chip surface in the experiments here, introduced an inhomogeneity in the DC field generated by electrodes E3 and E4. This DC field inhomogeneity is the cause of the larger residual uncanceled electric field of $F_\mathrm{u} = 104$~mV/cm in the current work. This residual uncanceled field is not randomly oriented. Its orientation at the location of the atoms, $\sim$\SI{300}{\micro\meter} above the chip surface, was determined from finite element calculations to be in the $z$ dimension -- as indicated by the central red vector crossing the distribution of atoms in Figure~\ref{fig:fields}. Similar finite element calculations of the frequency-dependent polarizations of the microwave dressing field and the evanescent microwave field from the CPW resonator indicate that at the position of the atoms both of these fields are also linearly polarized in the $z$ dimension, i.e., parallel to $\vec{F}_{\mathrm{u}}$.

\begin{figure}
    \centering
    \includegraphics[width=0.99\columnwidth]{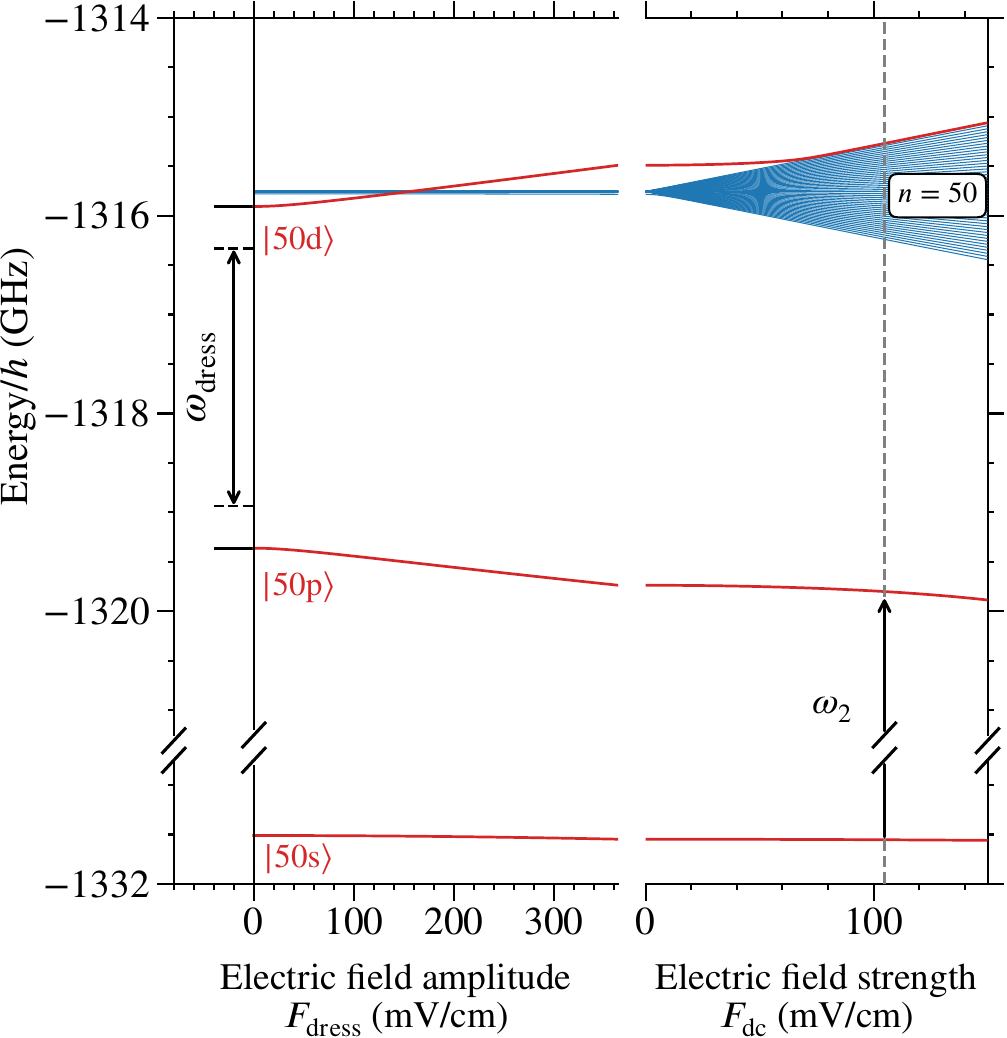}
    \caption{Overview of the calculated energy-level structure of the triplet Rydberg states in He with $n=50$ and $m_\ell=0$. The left part of the figure shows the AC Stark effect caused by an $\omega_\mathrm{dress} = 2\pi\times\SI{3.350}{GHz}$ microwave dressing field represented by the black vertical line, in the absence of DC electric fields. The right part shows the additional contribution from a DC offset electric field when $F_\mathrm{dress} = 367$~mV/cm, i.e., the upper limit of $F_\mathrm{dress}$ in the left panel. In the right part of the figure, the gray dashed vertical line indicates the residual DC electric field of $F_\mathrm{u}=104$~mV/cm encountered in the experiments, while the vertical arrow labeled $\omega_2$ represents the angular frequency of the second harmonic mode of the superconducting CPW resonator when $T_\mathrm{res}=\SI{3.66}{\kelvin}$. Note: For clarity only states with $q = 0$ are shown, but the calculations were carried out in a basis with $|q|\leq 3$.\label{fig:scheme}}
\end{figure}

\section{Microwave dressing scheme}\label{sec:dressingscheme}

The comparatively large quantum defects of the triplet $\sstate$ ($\delta_{\mathrm{50s}} \simeq 0.2966720$ \cite{drake_high_1999}) and  $\pstate$ ($\delta_{\mathrm{50p}} = 0.0683528$~\cite{drake_high_1999}) states in He means that they are energetically well-isolated from other higher-$\ell$ states with the same value of $n$. The single-photon electric-dipole allowed microwave transition between these states therefore remains strong in weak DC fields. For example, the transition moment is $d \simeq 2029\,e\,a_0$ in zero-field and remains $> 2000\,e\,a_0$ in DC fields up to $200$~mV/cm. The corresponding field-free transition frequency, $\omega_{\mathrm{50s, 50p}}$, lies close to $\omega_2$, with a detuning of $\Delta_{\mathrm{res}} = \omega_{\mathrm{50s, 50p}} - \omega_2 =  - 2\pi \times \SI{397.2}{MHz}$ when $T_\mathrm{res} = \SI{3.66}{\kelvin}$. Therefore, to resonantly couple the atoms to the second harmonic mode of the resonator using the $|50\mathrm{s}\rangle \rightarrow |50\mathrm{p}\rangle$ transition it must be tuned to lower frequency. Because the differential static electric dipole polarizability of the $|50\mathrm{s}\rangle$ and $|50\mathrm{p}\rangle$ states is positive, the DC Stark effect could, in principle, be used to induce the required frequency shift by applying an offset field of $F_{\mathrm{dc}} = 704$~mV/cm. However, in such a field, the $\pstate$ state is strongly polarized and increasingly sensitive to electric field noise, as it  approaches the $\ell$-mixed hydrogenic Rydberg-Stark states. Consequently, tuning the frequency of the $\sstate \rightarrow \pstate$ transition using a DC electric field in the region close to the superconducting chip surface is anticipated to lead to significant decoherence and spectral broadening.

The more effective way to tune the $\sstate \rightarrow \pstate$ transition into resonance with $\omega_2$ when $T_{\mathrm{res}} = 3.66$~K, is to use an off-resonant microwave dressing field to induce an AC Stark shift of the $\pstate$ state toward lower frequency. The required shift of the $\pstate$ state can be achieved by choosing the dressing field frequency to lie below the frequency of the $\pstate \rightarrow \dstate$ transition, with a detuning $\Delta_\mathrm{dress} = \omega_\mathrm{dress} - \omega_\mathrm{50s,\, 50d}$. In the presence of this microwave dressing field, the $\pstate$ state is AC Stark shifted to lower energy, while the $\dstate$ state is shifted to higher energy. This is seen on the left side of Figure \ref{fig:scheme}. In the experiments described here, the detuning of this dressing field from the field-free $\pstate \rightarrow \dstate$ transition frequency was chosen to be $\Delta_\mathrm{dress} = -2\pi \times \SI{102.77}{MHz}$.

In the region above the cryogenically cooled superconducting chip surface where the atoms interacted with the resonator, in addition to the dressing field, a residual uncanceled DC electric field of $F_\mathrm{u} = 104$~mV/cm was also present, as determined from the Stark shift in the spectrum in Figure~\ref{fig:fieldcancel}. This DC field introduces a $-2\pi\times \SI{7.927}{MHz}$ Stark shift of the $|50\mathrm{s}\rangle \rightarrow |50\mathrm{p}\rangle$ transition in the absence of the dressing field. To account for this in the calculations, the combined AC and DC Stark shifts of the $|50\mathrm{s}\rangle$ and $|50\mathrm{p}\rangle$ states were determined. This was achieved by considering the Hamiltonian of the Rydberg atom within the electric dipole approximation, such that,
\begin{equation}
	H(t) = H_0 + e F_{\mathrm{dc}}z + eF_\mathrm{dress}\cos(\omega_\mathrm{dress}t)\,z,\label{eq:Hdress}
\end{equation}
where $H_0$ is the field-free Hamiltonian, and the DC field $\vec{F}_{\mathrm{dc}} = [0, 0, F_{\mathrm{dc}}]$ and microwave dressing field  $\vec{F}_{\mathrm{dress}} = [0,0,F_{\mathrm{dress}} \cos (\omega_{\mathrm{dress}} t)]$ act in the $z$ dimension as discussed in Section~\ref{sec:calibration}.

Since the Hamiltonian in Equation~\ref{eq:Hdress} is periodic in time, it can be transformed into a time-independent Floquet Hamiltonian~\cite{ho_semiclassical_1983}. Following this transformation, the matrix representation of the Floquet Hamiltonian can be constructed in the atomic basis expanded by the number of Floquet modes, $|n\ell m_\ell q\rangle$,  where $\ell$ and $m_\ell$ are the orbital angular momentum and azimuthal quantum numbers of the Rydberg electron, respectively, and $q$ is the Floquet mode order, such that,
\begin{eqnarray}
\langle n'\ell'm_\ell'\,q' | H_F | n \ell m_\ell\,q \rangle &=&\dots\nonumber\\
& &\hspace{-2cm} \langle n'\ell'm_\ell'|H_0|n \ell m_\ell \rangle \delta_{q, q'} \dots\nonumber\\
& &\hspace{-1.6cm} +\,e F_{\mathrm{dc}} \langle n'\ell'm_\ell'|z|n l m_\ell \rangle \delta_{q, q'} \dots\nonumber \\
& &\hspace{-1.3cm} +\,\frac{e F_{\text{dress}}}{2} \langle n'\ell'm_\ell'|z|n \ell m_\ell \rangle \delta_{q \pm 1, q'} \dots \nonumber\\
& & \hspace{-1.0cm} +\,q \hbar \omega_{\text{dress}} \delta_{n, n'} \delta_{\ell, \ell'} \delta_{m_\ell, m_\ell'} \delta_{q, q'}.
\end{eqnarray}
In this expression, the first term on the right-hand side represents the field-free energy of the Rydberg states, i.e., $\langle n' \ell' m'_\ell|H_0 | n \ell m_\ell\rangle  = E_{n\ell} = -hcR_{\mathrm{He}}/(n-\delta_{n\ell})^2$. The second and third terms account for the interaction of the atom with the time-independent DC, and time-dependent microwave dressing fields, respectively. The final term represents the energy offset of the Floquet sidebands, which lie at integer multiples of the dressing field photon energy $q\hbar\omega_\mathrm{dress}$. The values of these matrix elements were determined by numerical integration using the Numerov method~\cite{zimmermanStarkStructureRydberg1979} and the quantum defects of the corresponding triplet Rydberg states in He~\cite{drake_high_1999}. The combined effects of the DC electric field, and microwave dressing field on the Rydberg energy-level structure were obtained from the eigenvalues of the Hamiltonian matrix. The corresponding eigenvectors were then used to determine the electric dipole moments for transitions between Rydberg states in the presence of the fields. Convergence was achieved in the eigenvalues, for the particular states of interest here with $n=50$ over the range of field strengths considered, using a basis set with $49\leq n \leq 51$, $|q_\mathrm{dress}| \leq 3$, and all allowed values of $\ell$.

An overview of the effect of DC electric fields up to 150~mV/cm on the AC Stark shifted $n=50$ Rydberg states in the presence of the strong microwave dressing field is shown on the right side of Figure~\ref{fig:scheme}. The uncanceled DC field of $F_{\mathrm{dc}} = F_\mathrm{u} = 104$~mV/cm encountered in the experiments is indicated by the dashed gray vertical line. The importance of accounting for this residual uncanceled field, to accurately calculate the effect of the microwave dressing field can be seen from the DC Stark shift of the microwave dressed $\pstate$ state. 

\begin{figure}
    \centering
    \includegraphics[width=0.95\columnwidth]{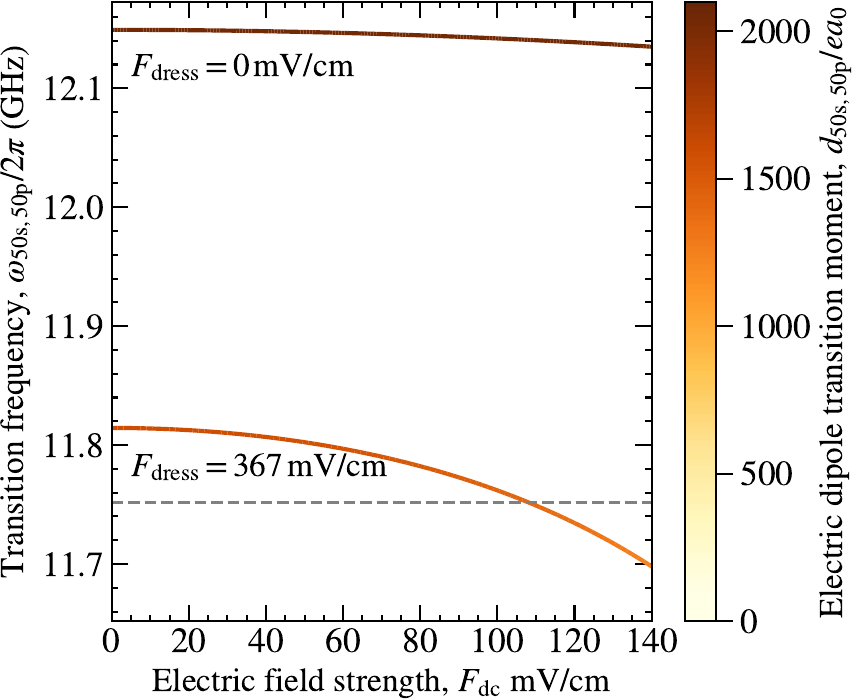}
    \caption{Calculated DC Stark shift and electric dipole transition moment of the $\sstate \rightarrow \pstate$ transition with ($F_\mathrm{dress} = 367$~mV/cm), and without ($F_\mathrm{dress} = 0$~V/cm) a $2\pi\times3.350$~GHz microwave dressing field applied.
     \label{fig:50s50ptransition}}
\end{figure}

A more detailed view of the calculated frequency shifts of the $\sstate \rightarrow \pstate$ transition can be seen in Figure~\ref{fig:50s50ptransition}. These data show that, in the presence of the residual DC electric field of $F_\mathrm{u} = 104$~mV/cm, a microwave dressing field amplitude of $367$~mV/cm is required to shift the $\sstate \rightarrow \pstate$ transition into resonance with $\omega_2$ when $T_\mathrm{res} = \SI{3.66}{\kelvin}$ (dashed horizontal line). Under these conditions the electric dipole moment of the $\sstate \rightarrow \pstate$ transition reduces from its field-free value $>2000\,e\,a_0$ as indicated by the color scale on the right side of the figure. However, it remains significant with a value of $d \sim 1500\,e a_0$ at the point where the lower curve, calculated in the presence of the microwave dressing field, crosses the dashed horizontal line.

\begin{figure}
    \centering
    \includegraphics[width=0.99\columnwidth]{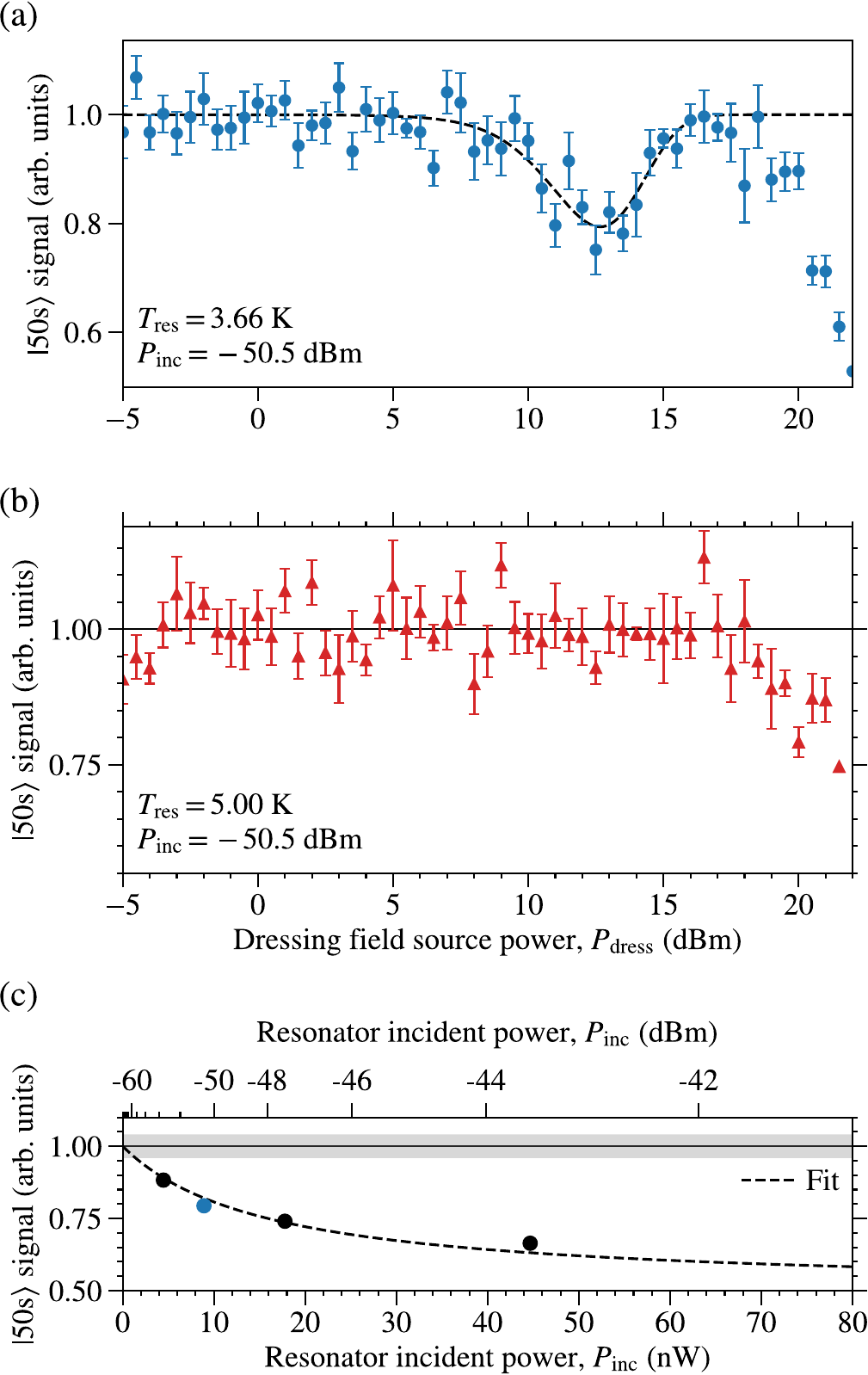}
    \caption{Dependence of the $\sstate$ population on the output power of the microwave source, $P_\mathrm{dress}$, used to generate the pulsed dressing field, when a 1-\SI{}{\micro\second}-duration $\omega_\mathrm{res} = 2\pi \times \SI{11.725}{GHz}$ field for which $P_{\mathrm{inc}} =-50.5$~dBm, propagated through the CPW. (a) $T_\mathrm{res} = \SI{3.66}{\kelvin}$ such that $\omega_2 = 2\pi \times \SI{11.725}{GHz}$, and (b) $T_\mathrm{res} = \SI{5.00}{\kelvin}$ so that $\omega_2$ was detuned by $-2\pi\times42.29$~MHz from the microwave field in the CPW. (c) Measured (filled points) dependence of the $\sstate$ population on the value of $P_{\mathrm{inc}}$ when $T_\mathrm{res} = \SI{3.66}{\kelvin}$ [as in panel~(a)] and $P_\mathrm{dress} = \SI{12}{dBm}$. The dashed curve in~(c) represents a fit to the experimental data (see text for details), while the half-width of the shaded gray band represents the standard deviation of the measured population.}
    \label{fig:5k}
\end{figure}

\section{Results}\label{sec:results}

To resonantly couple the single-photon $\sstate \rightarrow \pstate$ transition to the $\omega_2=2\pi\times11.751\,96$~GHz second harmonic mode of the superconducting CPW resonator when operated at $T_{\mathrm{res}} = 3.66$~K, it was necessary to determine the optimal output power from the microwave source, $P_{\mathrm{dress}}$, at which to drive the microwave antenna to generate a $\omega_\mathrm{dress} = 2\pi \times \SI{3.350}{GHz}$ microwave field with an amplitude of $367$~mV/cm above the resonator. This was found by injecting a 1-\SI{}{\micro\second}-duration pulse of microwave radiation into the resonator at its resonance frequency when the atoms were located above it, and monitoring the population of the $\sstate$ state. The peak power of this microwave pulse as it propagated through the CPW on the superconducting chip was $P_{\mathrm{inc}} = -50.5$~dBm. Since the microwave radiation at this frequency was far off resonance from $\omega_{\mathrm{50s, 50p}}$, it alone had no observable effect on the $\sstate$ population. With this weak microwave field in the resonator, a set of measurements were then made that involved the simultaneous application of the pulsed dressing field, while continuing to monitor the $\sstate$ population. The resulting data are presented in Figure~\ref{fig:5k}(a). From this figure it is seen that as $P_{\mathrm{dress}}$ is increased from -5 to $\sim+7$~dBm no significant change occurs in the population of the $\sstate$ state. However, as $P_{\mathrm{dress}}$ approaches 12~dBm, the $\sstate$ population reduces by $\sim0.2$ before returning again to 1 when $P_{\mathrm{dress}}$ reaches 16~dBm. This resonant depletion of the $\sstate$ population occurs when the $\sstate \rightarrow \pstate$ transition is AC Stark shifted into resonance with $\omega_2$, and the resulting coupling to the microwave field in the resonator causes a transfer of population to the $\pstate$ state.

From the results of the calculations shown in Figure~\ref{fig:50s50ptransition}, the strongest atom-resonator coupling is achieved when the AC Stark shifted $\sstate \rightarrow \pstate$ transition frequency first becomes equal to $\omega_2$. Consequently, the resonant depletion of the $\sstate$ population close to $P_{\mathrm{dress}}=12$~dBm in Figure~\ref{fig:5k}(a) indicates the value of $P_{\mathrm{dress}}$ at which this optimal dressing field is generated at the location of the atoms. The dashed black curve overlaid on these data represents a Gaussian function fit using least squares methods. From this, the ratio of the standard deviation in the dressing field amplitude across the distribution of atoms, $\sigma_{\mathrm{F}}$, to its mean value $F_\mathrm{dress}$, i.e., $\sigma_{\mathrm{F}}/F_\mathrm{dress}\simeq 0.2$, was determined. Comparison of this range of field amplitudes, with the results of the Floquet calculations indicate that this corresponds to a spectral FWHM in the frequency domain of $\simeq \SI{55}{MHz}$ when $F_\mathrm{dress} = 367$~mV/cm. Since this is larger than the Fourier transform limit of the 1-\SI{}{\micro\second}-duration pulsed field injected into the resonator, this suggests that the resonance widths encountered in the experiments when the $\sstate \rightarrow \pstate$ transition is AC Stark shifted into resonance with $\omega_2$, are strongly affected by the inhomogeneity of the microwave dressing field.

For values of $P_{\mathrm{dress}}\gtrsim17$~dBm in Figure~\ref{fig:5k}(a), further depletion of the $\sstate$ population is observed. This feature in the data, which occurs at microwave source powers 3 to 10 times larger than that required to tune the $\sstate\rightarrow\pstate$ transition into resonance with $\omega_2$, results from the complex multilevel character of the Rydberg states and the effect of the higher-$\ell$ states on the AC Stark shift of the $\pstate\rightarrow\dstate$ transition in strong fields. 

To confirm that the microwave dressed Rydberg atoms were resonantly coupled to the microwave field in the CPW resonator when $P_{\mathrm{dress}}=12$~dBm, and not, for example, the evanescent field of the CPW or microwave connectors on the PCB, a second similar measurement was made with the temperature of the superconducting chip increased to $T_{\mathrm{res}}=5.00$~K. This resulted in a $-2\pi\times42.29$~MHz frequency shift of $\omega_2$. In this situation the $2\pi \times \SI{11.75196}{GHz}$ microwave field for which $P_{\mathrm{inc}} = -50.5$~dBm, did not build up in the resonator upon pulsed injection through the CPW. Consequently, as seen from the corresponding data in Figure~\ref{fig:5k}(b), no resonant depletion of the $\sstate$ population is observed when $P_{\mathrm{dress}}=12$~dBm. 

With $T_{\mathrm{res}}=3.66$~K, and $P_{\mathrm{dress}}$ set to $12$~dBm, so that the AC Stark shift caused by the microwave dressing field ensured that the $\sstate \rightarrow \pstate$ transition was resonant with $\omega_2$, a further set of measurements were made of the dependence of the depletion of the $\sstate$ population on the value of $P_{\mathrm{inc}}$, i.e., the power of the microwave field in the CPW. These results are shown in Figure~\ref{fig:5k}(c). From these data (filled points) it is seen that as $P_{\mathrm{inc}}$ is reduced from -43.5~dBm (45~nW) to -53.5~dBm (4.5~nW) the depletion of the $\sstate$ population reduces from $\sim0.35$ to $\sim0.15$. Effects of saturation at the higher powers can be inferred from the dashed black curve representing the function~\cite{metcalf99a}
\begin{eqnarray}
P_{50\mathrm{s}} &=& 1 - \frac{s}{2(1+s)},
\end{eqnarray}
fit to the experimental data. In this expression the saturation parameter $s=B\,P_{\mathrm{inc}}$ on resonance, with the free fit parameter $B$ representing the inverse of the saturation power. Extrapolation of this fit function to values of $P_{\mathrm{inc}}$ approaching -60~dBm, indicates that for the \SI{1}{\micro\second} interaction time of the atoms with the resonator field in the experiments, microwave fields in the CPW with powers down to -58.5~dBm (1.4~nW) could be detected using the atoms, at frequencies resonant with $\omega_2$ and $\omega_{50\mathrm{s},50\mathrm{p}}$. This is the value of $P_{\mathrm{inc}}$ above which the depletion of the $\sstate$ population is greater than the standard deviation in the measured population, indicated by the shaded horizontal gray band. When $P_{\mathrm{inc}} = -43.5$~dBm (45~nW) on the right side of the figure, the microwave power is approximately twice that required to saturate the transition. 

 \begin{figure}
    \centering
    \includegraphics[width=0.8\linewidth]{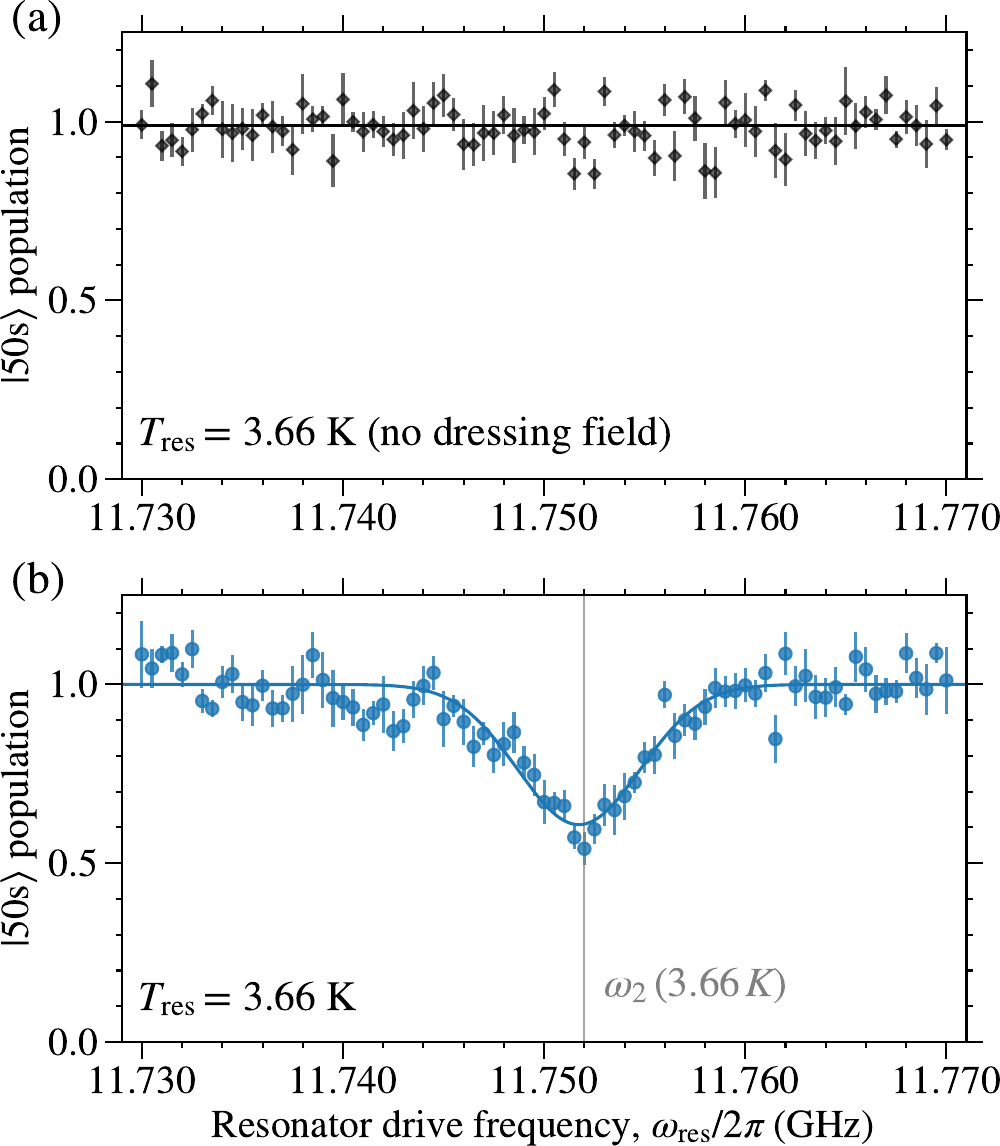}
    \caption{Microwave spectra of the $\sstate \rightarrow \pstate$ transition in He atoms coupled to the CPW resonator with the superconducting chip operated at $T_\mathrm{res} = \SI{3.66}{\kelvin}$. The data were recorded by varying the frequency of the microwave field injected into the CPW while monitoring the $\sstate$ population. The spectrum in panel (a) was recorded with the microwave dressing field off, so that the transition between the Rydberg states was far off resonant from $\omega_2$. That in panel (b) was recorded with the dressing field on and $P_{\mathrm{dress}}$ set to $12$~dBm to AC Stark shift the $\sstate \rightarrow \pstate$ into resonance with $\omega_2$.}
    \label{fig:spectra}
\end{figure}

To obtain frequency-domain information on the resonant coupling of the atoms to the resonator field in the presence of the strong microwave dressing field, spectra were recorded by varying the frequency of the microwave field in the CPW while monitoring the $\sstate$ population. These measurements were made for $T_{\mathrm{res}}= \SI{3.66}{\kelvin}$, so that $\omega_2 = 2\pi \times \SI{11.75196}{GHz}$ [as in Figure~\ref{fig:5k}(a)], and $P_{\mathrm{inc}}=-43.5$~dBm (45~nW). The results are displayed in Figure~\ref{fig:spectra}. Under these conditions, when the microwave dressing field was off, such that the $\sstate \rightarrow \pstate$ transition was off-resonance from the resonator, no depletion of the $\sstate$ population was observed as the frequency of the microwave field injected into it was adjusted [Figure~\ref{fig:spectra}(a)]. With the dressing field on and set to $P_{\mathrm{dress}}=12$~dBm, such that the AC Stark shifted $\sstate \rightarrow \pstate$ transition was resonant with $\omega_2$, depletion of the $\sstate$ population was observed at the frequency $\omega_2$ as seen in Figure~\ref{fig:spectra}(b). 

The centroid of the resonance in Figure \ref{fig:spectra}(b), determined by fitting a Gaussian function (continuous blue curve) to the experimental data using least squares methods, lies at $2\pi \times \SI{11.7517}{GHz}$ (continuous vertical gray line). This is in good agreement with the value of $\omega_2$ obtained from the microwave transmission measurements on the superconducting chip when $T_\mathrm{res} = \SI{3.66}{\kelvin}$ in Section \ref{sec:calibration}. The measured FWHM of the resonance, $\Delta \omega_{\mathrm{FWHM}} \simeq 2\pi \times \SI{7.1}{MHz}$, is greater than the Fourier transform limit of the \SI{1}{\micro\second}-duration microwave pulse injected into the resonator and exceeds its spectral FWHM, i.e., the resonator decay rate $\kappa = 2\pi\times \SI{2.97}{MHz}$, at this temperature. The corresponding spectral broadening is attributed to a combination of: (i) the inhomogeneity in the microwave dressing field in the region above the superconducting chip, which causes variations in the AC Stark shift of the $\sstate \rightarrow \pstate$ transition across the distribution of Rydberg atoms; (ii) the inhomogeneity in the DC field, to which this transition is sensitive because of its larger polarizability, for example, when compared to the single-color two-photon $|50\mathrm{s}\rangle \rightarrow |51\mathrm{s}\rangle$ transition in Figure \ref{fig:fieldcancel}; and (iii) saturation of the transition, which, for a saturation parameter of $s=2$, as determined from the data in Figure~\ref{fig:5k}(c), results in a contribution to the spectral width which is approximately equal to the Fourier transform limited FWHM of $2\pi\times1$~MHz~\cite{metcalf99a}.

The resonant behavior of the atom-resonator interaction in the presence of the microwave dressing field in Figure~\ref{fig:spectra} confirm that the calculated amplitude of $367$~mV/cm required to AC Stark shift the $\sstate \rightarrow \pstate$ transition into resonance with the resonator when operated at $\omega_2 = 2\pi \times \SI{11.75196}{GHz}$ is achieved when $P_{\mathrm{dress}}=12$~dBm. However, since the change in this AC Stark shift with field amplitude, i.e., the derivative of the Stark shift, in this field is $\sim2\pi\times1$~MHz/(mV/cm), a homogeneity of the dressing field of better than $3\times10^{-3}$ over the distribution of Rydberg atoms is required to achieve a spectral width of the transition that approaches the Fourier-transform limit of $2\pi\times1$~MHz. As seen from the diagram of the superconducting chip-holder and surrounding electrode structures in Figure~\ref{fig:fields}, these components, which strongly affect the spatial distribution of the microwave dressing field emanating from the antenna, have size scales on the order of 10~mm. Since the spatially extended (6-mm-long in the $y$ dimension by \SI{100}{\micro\meter} wide in the $x$ and $z$ dimensions) distribution of atoms moving $\sim$\SI{300}{\micro\meter} above the superconducting chip surface has a size between $10^{-2}$ and $10^{-1}$ of that of the electrode structures, a field homogeneity $<3\times10^{-3}$ across the bunch of atoms is challenging to achieve. Indeed, even the homogeneity of $10^{-2}$, required to ensure that the range of AC Stark shifts across the bunch of atoms remains below the $2\pi\times2.97$~MHz spectral FWHM of the resonator mode, places stringent requirements on the apparatus. Based on these considerations we conclude that it is the inhomogeneity in the microwave dressing field that dominants the spectral broadening in the measurements. 

\section{Discussion}\label{sec:discuss}

The inhomogeneity of the microwave dressing field across the spatial distribution of atom positions that contributes to the broadening of the resonance in Figure~\ref{fig:spectra}(b) also gives rise to dephasing of the atom-resonator interaction. This dephasing is a result of the corresponding distribution of frequency detunings from the resonator resonance frequency, and to a lesser extent the distribution of resonant Rabi frequencies. The use of a fast moving beam of Rydberg atoms in the experiments here offers the advantage of easily separating the Rydberg state laser photoexcitation region, from the region in which the atoms interact with the superconducting circuit, and the detection region. However, it has the disadvantage of yielding comparatively long (typically 2 -- 6~mm long) bunches of excited Rydberg atoms that move up to 2~mm within a \SI{1}{\micro\second} atom-resonator interaction time. In the future, narrower spectral features and reduced dephasing can be achieved by laser slowing and cooling the metastable He atoms~\cite{bardouMagnetoOpticalTrappingMetastable1992, rooijakkersLaserDecelerationTrapping1997} and ultimately positioning them precisely above the CPW resonator, for example, using optical tweezers. With cold atoms localized in this way, and a resonator fabricated with a resonance frequency closer to the field-free $\sstate \rightarrow \pstate$ transition frequency to reduce the required AC Stark shift, the inhomogeneity of the microwave dressing field will have a reduced effect on the measurements. Close to the CPW resonator, when the atom-surface distance is on the same order of magnitude as the features in the superconducting circuit, i.e., \SI{10}{\micro\meter}, it will be important to treat more completely the possibility of a non-zero angle between the polarizations of the dressing field and the resonator field. This can be achieved by expanding the Floquet calculations to include a time-varying component polarized in the $x$ or $y$ dimensions. 

From the data in Figure~\ref{fig:5k}(c), when $P_\mathrm{inc} = \SI{-53.5}{dBm}$ the depletion of the $\sstate$ population after the \SI{1}{\micro\second} interaction time of the atoms with the resonator is $\sim0.1$. This indicates that the resonant Rabi frequency associated with the atom-resonator interaction under these conditions is approximately $2\pi\times100$~kHz. The photon occupation number of the resonator when $P_\mathrm{inc} = \SI{-53.5}{dBm}$ can be estimated from the circulating power in it, $P_{\mathrm{circ}}$, where~\cite{sage11a}
\begin{equation}\label{eq:pinc}
	P_\mathrm{circ} = P_\mathrm{inc} Q \frac{10^{-\frac{L_\mathrm{ins}}{20}}}{m\pi},
\end{equation}
\noindent
with $m=2$ and $Q=3960$ being the mode number and quality factor of the resonator, respectively, and the insertion loss $L_\mathrm{ins} = \SI{26.96}{dB}$ (see Section \ref{sec:calibration}). Under these conditions $P_{\mathrm{circ}}$ is estimated to be $\sim$\SI{0.1}{\micro\watt} which corresponds to a photon occupation number in the second harmonic mode of $N \sim 10^6$. Since the Rabi frequency associated with the resonant interaction of the atoms with the resonator is dependent on $\sqrt{N}$, the single-photon Rabi frequency, i.e., the Rabi frequency when $N=1$, in the current experiments can be estimated to be $2\pi\times100$~Hz. 

In the future this single-photon Rabi frequency can be increased by improving the homogeneity of the DC electric fields generated above the CPW resonator. This can be achieved by biasing the antenna to the same offset potential as the split electrodes E3 and E4. A further increase will be achieved by fabricating a resonator with a resonance frequency closer to the field-free Rydberg-Rydberg transition frequency, and thus reducing the AC Stark shift required to tune the transition into resonance with it. This will mean that the electric dipole transition moment between these states with the dressing field applied will increase and approach its field-free value. 

However, the most significant increase in atom-resonator coupling will be seen by positioning the atoms closer to the resonator and the surface of the superconducting chip. In the current experiments the atoms are located $\sim$\SI{300}{\micro\meter} above the chip surface when they interact with the resonator~\cite{walkerElectrometrySingleResonator2022}. If this distance is reduced to between $30$ and \SI{50}{\micro\meter}, single-photon Rabi frequencies on the order of $2\pi\times1$~MHz are expected to be achievable because of the corresponding increase in the amplitude of the evanescent field close to the resonator.

\section{Conclusion}\label{sec:conc}

The results presented here represent an important step toward reaching the single-photon strong-coupling regime in hybrid microwave cavity QED with Rydberg atoms coupled to CPW resonators in superconducting circuits. They demonstrate control of the Rydberg-Rydberg transition frequency at this hybrid interface using a strong off-resonant microwave dressing field, and represent the first demonstration of coupling a strong ($d \sim 1500\,\mathrm{ea}_0$) single-photon microwave transition between Rydberg states to a superconducting CPW resonator. Since this transition dipole moment is $\sim30$ times larger than in any previous related experiments, the circulating power in the resonator could be reduced by almost three orders of magnitude over that employed previously. Although the experiments were performed with a microwave field injected into the resonator to ensure an observable population transfer between the Rydberg states in the \SI{1}{\micro\second} interaction time of the atoms with it, consideration of the circulating power in the resonator and corresponding photon occupation number, allowed the single-photon Rabi frequency to be estimated as $2\pi\times100$~Hz. With future upgrades to the apparatus to better localize the atoms, and move them closer to the surface of the superconducting chip, limitations arising from the homogeneity of the microwave dressing field will be overcome to allow this type of microwave dressed atom-circuit interface to be exploited for optical manipulation of the microwave-photon number in the resonator, or ultimately the quantum state of a superconducting qubit in an interaction mediated by the atoms. 

\begin{acknowledgments}
This work was supported the Engineering and Physical Sciences Research Council (EPSRC) through Grant No. EP/Y022688/1, the EPSRC Centre for Doctoral Training in Delivering Quantum Technologies (Grant No. EP/S021582/1), and the UCL EPSRC Doctoral Training Partnership (Grant No. EP/W524335/1).
\end{acknowledgments}

\bibliography{hybrid_ac_shifted_50s_50p_transition}

\end{document}